\documentclass[conference]{IEEEtran}
\IEEEoverridecommandlockouts
\usepackage{cite}
\usepackage{amsmath,amssymb,amsfonts}
\usepackage{algorithmic}
\usepackage{graphicx}
\usepackage{textcomp}
\usepackage{hyperref}
\usepackage{amsmath,amsthm}
\usepackage{braket}
\usepackage[table,xcdraw]{xcolor}
\usepackage{tikz}
\usetikzlibrary{quantikz2}
\def\BibTeX{{\rm B\kern-.05em{\sc i\kern-.025em b}\kern-.08em
    T\kern-.1667em\lower.7ex\hbox{E}\kern-.125emX}}
\usepackage{etoolbox}
\makeatletter
\patchcmd{\@makecaption}
  {\scshape}
  {}
  {}
  {}
\makeatother
\begin{document}

\title{Validation of Quantum Elliptic Curve Point Addition Circuits

{\normalsize

\today}
}
\author{\IEEEauthorblockN{Francis P. Papa}
\IEEEauthorblockA{\textit{Google} \\
Venice, CA 90291, USA \\
effpapa@google.com}
}

\maketitle

\begin{abstract}
Specific quantum algorithms exist to---in theory---break elliptic curve cryptographic protocols. Implementing these algorithms requires designing quantum circuits that perform elliptic curve arithmetic. To accurately judge a cryptographic protocol's resistance against future quantum computers, researchers figure out minimal resource-count circuits for performing these operations while still being correct. To assure the correctness of a circuit, it is integral to restore all ancilla qubits used to their original states. Failure to do so could result in decoherence of the computation's final result. Through rigorous classical simulation and unit testing, I surfaced four inconsistencies in the state-of-the-art quantum circuit for elliptic curve point addition where the circuit diagram states the qubits are returned in the original ($\ket{0}$) state, but the intermediate values are not uncomputed. I provide fixes to the circuit without increasing the leading-order gate cost.
\end{abstract}

\section{Introduction}
Elliptic curve cryptography is subject to attack by quantum computers of sufficient size and quality inspired by Shor's original algorithm for the discrete logarithm \cite{shor}. The quantum algorithm centers around phase estimation of elliptic curve point addition. A recent optimized construction of the quantum circuit for elliptic curve point addition operates in six steps and requires $5n+5$ ancilla qubits allocated for flags and intermediate calculations \cite{ecc-psi}. The circuit is intended to be exact for the point addition operation. It indicates the ancillae to enter in the $\ket{0}$ state, indicates them to end in the $\ket{0}$ state, and intentionally attempts to clear them in the circuit. However, in a few rare cases, the ancilla qubits are not actually returned to their original state.

Qualtran is a framework for expressing and analyzing quantum algorithms \cite{qualtran-paper}. The framework allows users to quickly simulate quantum operations in purely classical states efficiently. Leveraging this, users are able to introduce circuits to the library built on correctness from the ground up. In this manuscript, I report an implementation of the Litinski \cite{ecc-psi} elliptic curve arithmetic primitives in the Qualtran quantum programming language. I validate the constructions by classical simulation of their action on computational basis states, and verify the resource costs of the constructions---particularly the count of Toffoli gates.

\section{Methodology}
I implemented Shor's algorithm for breaking elliptic curve private keys in the Qualtran framework following the circuit schematics from Ref.~\cite{ecc-psi}. Using ordinary arithmetic primitive implementations from the Qualtran standard library of subroutines I created a hierarchical definition of the circuit composed of subroutines from the lowest level—bitwise addition through basic quantum gates—to the top level of quantum phase estimation. This allowed me to validate the correctness of each circuit and the gate counts reported in the papers describing them.

To validate correctness, I took advantage of several key abstractions provided by the Qualtran framework. Each subroutine was transcribed into a Bloq; Bloqs are objects that represent a quantum subroutine. The Bloq object has a signature to keep track of the input and output registers, member methods that emulate the classical function of the circuit being translated, circuit compositions in the form of directed acyclic graphs, and an exact inventory of the required subroutines or Bloqs. When translating the circuit schematics into a Bloq decomposition, the developer can also introduce and manage any ancilla qubits required. To account for the additional care that must be taken to reset ancilla qubits to their input states, errors will be generated if the qubits are not returned to their original state. The construction uses integers in the Montgomery form \cite{montgomery-mult} as an optimization, so I also implemented a new quantum data type annotation for quantum unsigned integers in Montgomery form. This class includes member functions that translate classical unsigned integers to and from Montgomery form and serves as a reference for computation of the Montgomery product and inverse according to Gouzien et al. \cite{montgomery}.

Validation of the elliptic curve arithmetic circuits was accomplished through three types of unit tests: validity checks, classical simulation tests, and Toffoli cost tests.

\subsection{Validity Checks}
The Qualtran framework provides testing functions to assert the validity of Bloq decompositions. These tests confirm that the circuit DAG is valid (Bloqs have the correct number of inputs/outputs, all nodes on the graph connect to something, nodes only have one connection, node bitsizes match connected registers, and data type annotations match). Additionally, the validity unit tests check that the Bloq counts gathered from explicitly counting the Bloqs used in the decomposition match the expected, reference value.

\subsection{Classical Simulation Tests}
The Qualtran framework also implements classical simulation when applicable. It does this by piping classical data through the circuit and manipulating it in Bloqs using the corresponding classical function. Correctness can be tested by verifying that the classical data as computed by the circuit matches the value expected by ordinary classical computation of the reference value. Checks can be propagated hierarchically and efficiently by assuming the correctness of low-level Bloqs used in the composition of a higher-level Bloq. Unit tests will fail if one tries to free qubits that are not in the $\ket{0}$ state. For elliptic curve point addition, I designed a Bloq that decomposes into the six steps diagrammed by Ref. \cite{ecc-psi} and designed Bloqs for each of these steps which decomposed further into the subroutines required for each step. In the unit tests, I utilized a subset of points from the p1707 curve described in Ref. \cite{ecc-curve}. In each test, I chose 12 of these points, converted them into Montgomery form, piped them through the circuits, and compared the results to the expected reference values.

\subsection{Toffoli Cost Tests}
Lastly, Qualtran allows one to represent resource costs—in particular, qubit and gate counts—using symbolic classical parameters. One can hierarchically test the correctness of important gate counts when using symbolic bitsizes. In this case, I validate the Toffoli counts of the subroutines against the reference expressions given in Ref.~\cite{ecc-psi}, Figs. 5 and 8 to leading order. 

\section{Results}
Via these three categories of checks, I isolated four inconsistencies in the elliptic curve point addition circuits which cause ancilla qubits not to be cleared or to be incorrectly cleared in steps 2, 5, and 6. I propose modifications to the circuits which properly clear the ancilla flag bits in these cases without meaningfully increasing the number of Toffoli gates and only increasing the number of ancilla qubits by one. The following discussion assumes a familiarity with the construction and notation given in Ref. \cite{ecc-psi}, including the circuits in Figs. 10-12.

\subsection{Step 2}
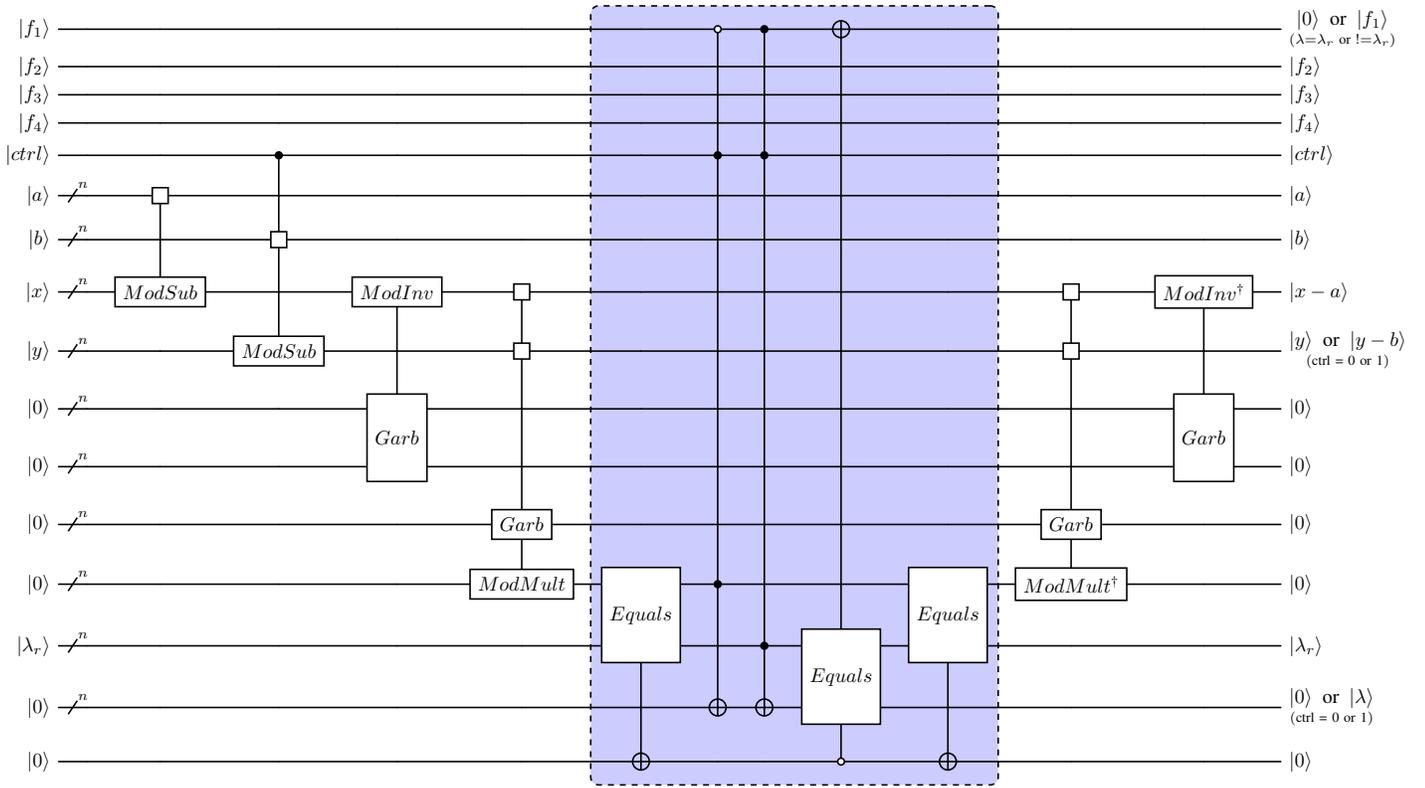
\begin{figure*}[htbp]
    \centering
    \scalebox{.75}{
    \begin{quantikz}
\lstick{$\ket{f_1}$} &&&&&&\gategroup[wires=16,steps=5,style={dashed,rounded
corners,fill=blue!20, inner
xsep=2pt},background]{}& \ocontrol{}\wire[d][4]{q} & \control{}\wire[d][4]{q} & \targ{}\wire[d][13]{q} &&&& \rstick{$\underset{(\lambda = \lambda_r \textrm{ or } != \lambda_r)}{\ket{0} \textrm{ or } \ket{f_1}}$} \\
\lstick{$\ket{f_2}$} &&&&&&&&&&&&& \rstick{$\ket{f_2}$} \\
\lstick{$\ket{f_3}$} &&&&&&&&&&&&& \rstick{$\ket{f_3}$} \\
\lstick{$\ket{f_4}$} &&&&&&&&&&&&& \rstick{$\ket{f_4}$} \\
\lstick{$\ket{ctrl}$} &&& \ctrl{2} &&&& \control{}\wire[d][8]{q} & \control{}\wire[d][9]{q} &&&&& \rstick{$\ket{ctrl}$} \\
\lstick{$\ket{a}$} & \qwbundle{n} & \gate{}\wire[d][2]{q} &&&&&&&&&&& \rstick{$\ket{a}$} \\
\lstick{$\ket{b}$} & \qwbundle{n} && \gate{}\wire[d][2]{q} &&&&&&&&&& \rstick{$\ket{b}$} \\
\lstick{$\ket{x}$} & \qwbundle{n} & \gate{ModSub} && \gate{ModInv}\wire[d][2]{q} & \gate{}\wire[d][1]{q} &&&&&& \gate{}\wire[d][1]{q} & \gate{ModInv\textsuperscript{\textdagger}}\wire[d][2]{q} & \rstick{$\ket{x-a}$} \\
\lstick{$\ket{y}$} & \qwbundle{n} && \gate{ModSub} && \gate{}\wire[d][3]{q} &&&&&& \gate{}\wire[d][3]{q} && \rstick{$\underset{(\textrm{ctrl = 0 or 1})}{\ket{y} \textrm{ or } \ket{y-b}}$} \\
\lstick{$\ket{0}$} & \qwbundle{n} &&& \gate[2]{Garb} &&&&&&&& \gate[2]{Garb} & \rstick{$\ket{0}$} \\
\lstick{$\ket{0}$} & \qwbundle{n} &&&&&&&&&&&& \rstick{$\ket{0}$} \\
\lstick{$\ket{0}$} & \qwbundle{n} &&&& \gate{Garb}\wire[d][1]{q} &&&&&& \gate{Garb}\wire[d][1]{q} && \rstick{$\ket{0}$} \\
\lstick{$\ket{0}$} & \qwbundle{n} &&&& \gate{ModMult} & \gate[2]{Equals}\wire[d][3]{q} & \control{}\wire[d][2]{q} &&& \gate[2]{Equals}\wire[d][3]{q} & \gate{ModMult\textsuperscript{\textdagger}} && \rstick{$\ket{0}$} \\
\lstick{$\ket{\lambda_r}$} & \qwbundle{n} &&&&&&& \control{}\wire[d][1]{q} & \gate[2]{Equals}\wire[d][2]{q} &&&& \rstick{$\ket{\lambda_r}$} \\
\lstick{$\ket{0}$} & \qwbundle{n} &&&&&& \targ{} & \targ{} &&&&& \rstick{$\underset{(\textrm{ctrl = 0 or 1})}{\ket{0} \textrm{ or } \ket{\lambda}}$} \\
\lstick{$\ket{0}$} &&&&&& \targ{} &&& \ocontrol{} & \targ{} &&& \rstick{$\ket{0}$}
\end{quantikz}
}
    \caption{Corrected circuit for step 2 of the elliptic curve point addition circuit; the changed part is highlighted in blue. In place of a single Equals gate, two Equals gates, a controlled-Equals gate, and an ancilla qubit are used to only clear the $\ket{f_1}$ flag when the computed $\lambda$ is not used and $\lambda_r$ is used instead.}
    \label{step-2}
\end{figure*}

In step 2, the circuit computes $\lambda$ into the ancilla register when $\ket{f_1} = \ket{0}$ and copies $\lambda_r$ into the ancilla register when $\ket{f_1} = \ket{1}$. The circuit wrongfully assumes that the calculated $\ket{\lambda} \neq \ket{\lambda_r}$ and uses this equivalence condition to clear the $f_1$ flag. In rare cases this assumption is false and causes $f_1$ to be left in the $\ket{1}$ state.

To correct this, I used one additional clear ancilla bit which is flipped on when the computed $\ket{\lambda}=\ket{\lambda_r}$; it is later uncomputed and freed. This change can be seen in Fig. \ref{step-2} highlighted in blue. The $\textit{Equals}$ operation that clears $\ket{f_1}$ is controlled on the new ancilla bit being $\ket{0}$ so that in the rare case described above, clearing of $\ket{f_1}$ is either left to step 6 or $\ket{f_1}$ is not cleared at all (because it is already $\ket{0}$). The original intention of this $\textit{Equals}$ operation on $\ket{f_1}$ was to detect the case that $\lambda_r$ is copied into the ancilla $\ket{\lambda}$ register. If $\ket{f_1}=\ket{1}$ and the computation is point doubling, then $\ket{f_1}$ wouldn't be cleared in step 6; it is only cleared in step 6 if $\ket{f_1}=\ket{f_2}=\ket{1}$ which would cause $\ket{(x_r,y_r)}=\ket{(0, 0)}$. In this case, because the computed $\lambda$ value is in the $\ket{\lambda}$ register, the computation is not point doubling and so $\ket{f_1}$ should not be cleared.

\subsection{Step 5}
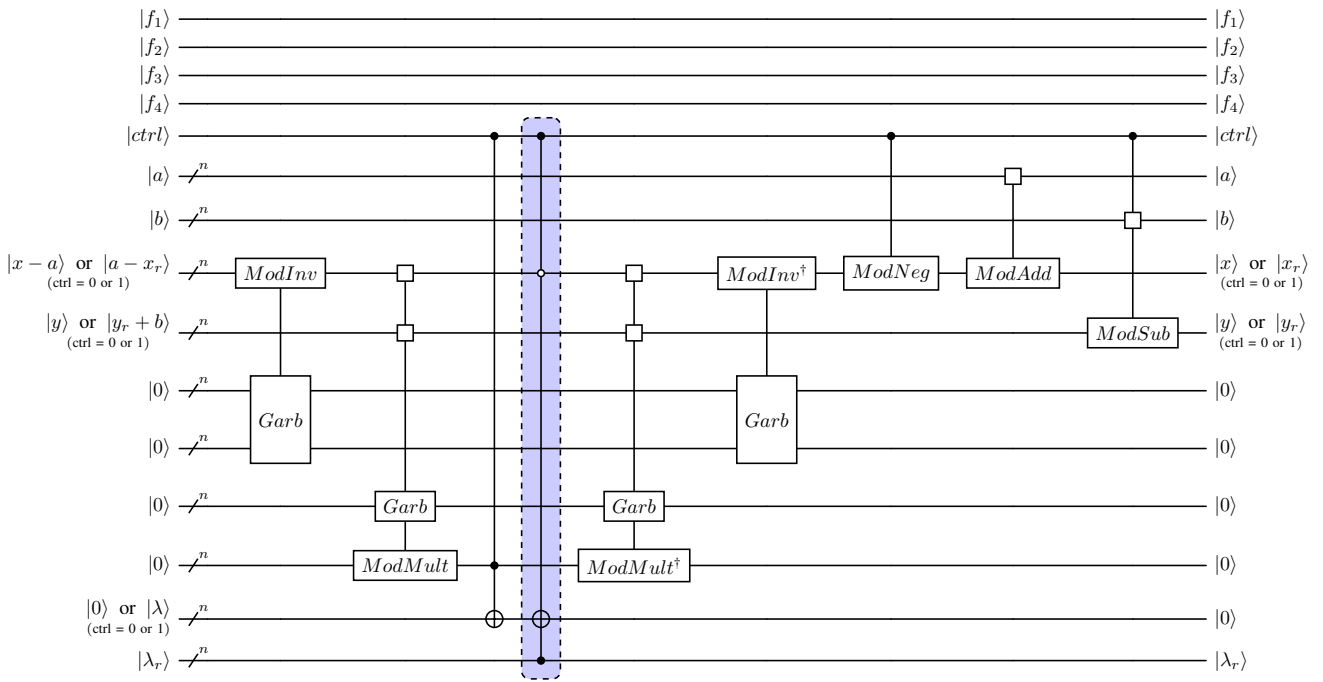
\begin{figure*}[htbp]
    \centering
    \scalebox{.75}{
    \begin{quantikz}
\lstick{$\ket{f_1}$} &&&&&&&&&&& \rstick{$\ket{f_1}$} \\
\lstick{$\ket{f_2}$} &&&&&&&&&&& \rstick{$\ket{f_2}$} \\
\lstick{$\ket{f_3}$} &&&&&&&&&&& \rstick{$\ket{f_3}$} \\
\lstick{$\ket{f_4}$} &&&&&&&&&&& \rstick{$\ket{f_4}$} \\
\lstick{$\ket{ctrl}$} &&&& \control{}\wire[d][8]{q} & \control{}\wire[d][3]{q}\gategroup[wires=11,steps=1,style={dashed,rounded
corners,fill=blue!20, inner
xsep=2pt},background]{} &&& \control{}\wire[d][3]{q} && \control{}\wire[d][2]{q} & \rstick{$\ket{ctrl}$} \\
\lstick{$\ket{a}$} & \qwbundle{n} &&&&&&&& \gate{}\wire[d][2]{q} && \rstick{$\ket{a}$} \\
\lstick{$\ket{b}$} & \qwbundle{n} &&&&&&&&& \gate{}\wire[d][2]{q} & \rstick{$\ket{b}$} \\
\lstick{$\underset{(\textrm{ctrl = 0 or 1})}{\ket{x-a} \textrm{ or } \ket{a-x_r}}$} & \qwbundle{n} & \gate{ModInv}\wire[d][2]{q} & \gate{}\wire[d][1]{q} && \ocontrol{}\wire[d][6]{q} & \gate{}\wire[d][1]{q} & \gate{ModInv\textsuperscript{\textdagger}}\wire[d][2]{q} & \gate{ModNeg} & \gate{ModAdd} && \rstick{$\underset{(\textrm{ctrl = 0 or 1})}{\ket{x} \textrm{ or } \ket{x_r}}$} \\
\lstick{$\underset{(\textrm{ctrl = 0 or 1})}{\ket{y} \textrm{ or } \ket{y_r+b}}$} & \qwbundle{n} && \gate{}\wire[d][3]{q} &&& \gate{}\wire[d][3]{q} &&&& \gate{ModSub} & \rstick{$\underset{(\textrm{ctrl = 0 or 1})}{\ket{y} \textrm{ or } \ket{y_r}}$} \\
\lstick{$\ket{0}$} & \qwbundle{n} & \gate[2]{Garb} &&&&& \gate[2]{Garb} &&&& \rstick{$\ket{0}$} \\
\lstick{$\ket{0}$} & \qwbundle{n} &&&&&&&&&& \rstick{$\ket{0}$} \\
\lstick{$\ket{0}$} & \qwbundle{n} && \gate{Garb}\wire[d][1]{q} &&& \gate{Garb}\wire[d][1]{q} &&&&& \rstick{$\ket{0}$} \\
\lstick{$\ket{0}$} & \qwbundle{n} && \gate{ModMult} & \control{}\wire[d][1]{q} && \gate{ModMult\textsuperscript{\textdagger}} &&&&& \rstick{$\ket{0}$} \\
\lstick{$\underset{(\textrm{ctrl = 0 or 1})}{\ket{0} \textrm{ or } \ket{\lambda}}$} & \qwbundle{n} &&&\targ{} & \targ{}\wire[d][1]{q} &&&&&& \rstick{$\ket{0}$} \\
\lstick{$\ket{\lambda_r}$} & \qwbundle{n} &&&& \control{} &&&&&& \rstick{$\ket{\lambda_r}$}
\end{quantikz}
}
    \caption{Corrected circuit for step 5 of the elliptic curve point addition circuit; the changed part is highlighted in blue. A $2n$-controlled Toffoli gate is used to  clear the $\ket{\lambda}$ register using the $\ket{\lambda_r}$ register when the modular inverse of $a-x_r$ is $0$.}
    \label{step-5}
\end{figure*}

In step 5, the circuit clears the $\ket{\lambda}$ register by uncomputing $\lambda$ from $a-x_r$ and $y_r+b$. However, in rare cases $a-x_r =0$, meaning that the modular inverse is undefined. This leaves $\ket{\lambda}$ in some nonzero state depending on your implementation of $\textrm{ModInv}(\ket{0})$ (in this case $\textrm{ModInv}(\ket{0})=\ket{0}$ so the register is unchanged).

First, I will prove that this inconsistency arises only during point doubling.

\begin{proof}
  Let $P_1,P_2$ be points on some elliptic curve such that $P_1 \neq P_2$. Additionally, $a-x_r=0$.
  From Eq. 2 in \cite{ecc-psi}, it is known that
  \begin{align*}
    \lambda &= \frac{y-b}{x-a} = \frac{y_r+b}{a-x_r} \mod p \\
    x-a     &= a-x_r =0 \mod p\\
    x &=a = x_r \mod p
  \end{align*}
  Because $P_1 \neq P_2$ and $a=x$, $y \neq b$.
  Again, from Eq. 2, it is known that
  \begin{align*}
    x_r &= \lambda^2 - x - a \mod p\\
    x_r - a &= \lambda^2 - x -2a \mod p\\
    0 &= \lambda^2 - x -2a \mod p&&\text{(because $x_r-a=0$)}\\
    0 &= (\frac{y-b}{x-a})^2 -x -2a \mod p&&\text{(substituting $\lambda$)}\\
    x+2a &=(\frac{y-b}{x-a})^2 \mod p\\
    (x-a)^2(x+2a) &= (y-b)^2 \mod p\\
    0 &= (y-b)^2 \mod p&&\text{(because $x-a=0$)} \\
    0 &= y-b \mod p\\
    b &= y \mod p
  \end{align*}
  However, it has already been established that $b \neq y$ because that would indicate point doubling. Therefore, $P_1=P_2$ and the operation must be point doubling.
\end{proof}
Thus, the computation is point doubling and $\ket{\lambda}=\ket{\lambda_r}=\ket{\frac{3a^2+c_1}{2b}}$. Therefore, it is sufficient to do $\ket{\lambda}\oplus\ket{\lambda_r}$ controlled on $\ket{ctrl} =\ket{1}$ and $\ket{a-x_r}=\ket{0}$ after the other attempt to clear the register. The change is highlighted in blue in Fig. \ref{step-5}.

\subsection{Step 6}
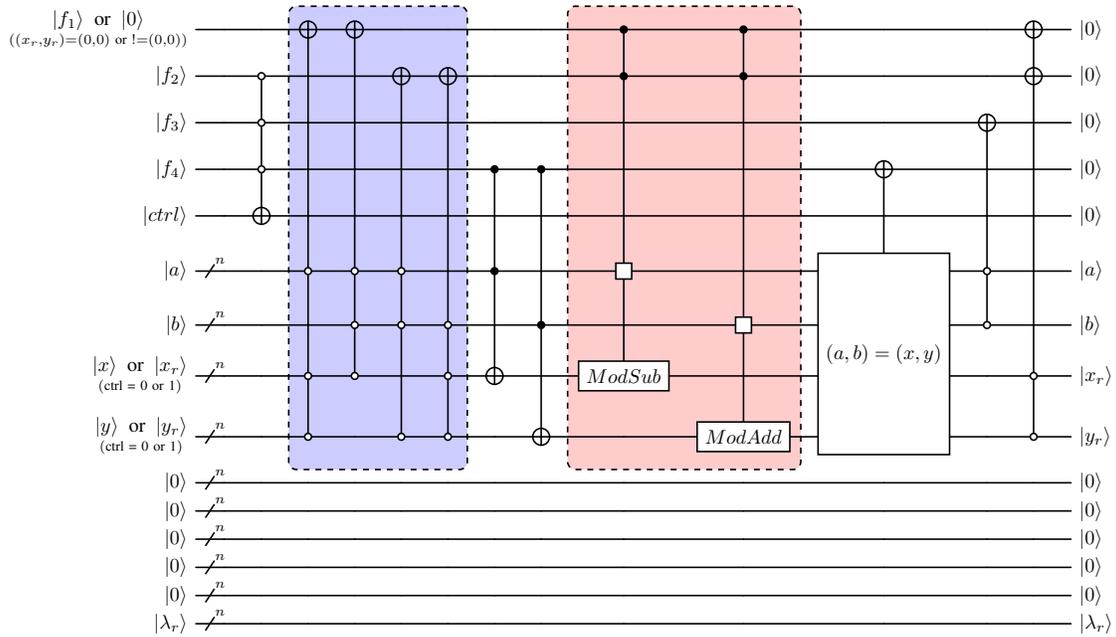
\begin{figure*}[htbp]
    \centering
    \scalebox{.75}{
    \begin{quantikz}
\lstick{$\underset{((x_r,y_r) = (0,0) \textrm{ or } != (0,0))}{\ket{f_1} \textrm{ or } \ket{0}}$} &&&\targ{}\wire[d][5]{q}\gategroup[wires=9,steps=4,style={dashed,rounded
corners,fill=blue!20, inner
xsep=2pt},background]{} & \targ{}\wire[d][5]{q}&&&&& \control{}\wire[d][1]{q}\gategroup[wires=9,steps=2,style={dashed,rounded
corners,fill=red!20, inner
xsep=2pt},background]{} & \control{}\wire[d][1]{q} &&& \targ{}\wire[d][1]{q} & \rstick{$\ket{0}$} \\
\lstick{$\ket{f_2}$} && \ocontrol{}\wire[d][1]{q} &&&\targ{}\wire[d][4]{q} & \targ{}\wire[d][5]{q}&&& \control{}\wire[d][4]{q} & \control{}\wire[d][5]{q} &&& \targ{}\wire[d][6]{q} & \rstick{$\ket{0}$} \\
\lstick{$\ket{f_3}$} && \ocontrol{}\wire[d][1]{q} &&&&&&&&&& \targ{}\wire[d][3]{q} && \rstick{$\ket{0}$} \\
\lstick{$\ket{f_4}$} && \ocontrol{}\wire[d][1]{q} &&&&& \control{}\wire[d][2]{q} & \control{}\wire[d][3]{q} &&& \targ{}\wire[d][2]{q} &&& \rstick{$\ket{0}$} \\
\lstick{$\ket{ctrl}$} && \targ{} &&&&&&&&&&&& \rstick{$\ket{0}$} \\
\lstick{$\ket{a}$} & \qwbundle{n} &&\ocontrol{}\wire[d][2]{q}&\ocontrol{}\wire[d][1]{q}& \ocontrol{}\wire[d][1]{q} && \control{}\wire[d][2]{q} && \gate{}\wire[d][2]{q} && \gate[4]{(a,b) = (x,y)} & \ocontrol{}\wire[d][1]{q} && \rstick{$\ket{a}$} \\
\lstick{$\ket{b}$} & \qwbundle{n} &&& \ocontrol{}\wire[d][1]{q} & \ocontrol{}\wire[d][2]{q} &\ocontrol{}\wire[d][1]{q}&& \control{}\wire[d][2]{q} && \gate{}\wire[d][2]{q} && \ocontrol{} && \rstick{$\ket{b}$} \\
\lstick{$\underset{(\textrm{ctrl = 0 or 1})}{\ket{x} \textrm{ or } \ket{x_r}}$} & \qwbundle{n} &&\ocontrol{}\wire[d][1]{q}&\ocontrol{}&& \ocontrol{}\wire[d][1]{q}& \targ{} && \gate{ModSub} &&&& \ocontrol{}\wire[d][1]{q} & \rstick{$\ket{x_r}$} \\
\lstick{$\underset{(\textrm{ctrl = 0 or 1})}{\ket{y} \textrm{ or } \ket{y_r}}$} & \qwbundle{n} &&\ocontrol{}&& \ocontrol{} &\ocontrol{}&& \targ{} && \gate{ModAdd} &&& \ocontrol{} & \rstick{$\ket{y_r}$} \\
\lstick{$\ket{0}$} & \qwbundle{n} &&&&&&&&&&&&& \rstick{$\ket{0}$} \\
\lstick{$\ket{0}$} & \qwbundle{n} &&&&&&&&&&&&& \rstick{$\ket{0}$} \\
\lstick{$\ket{0}$} & \qwbundle{n} &&&&&&&&&&&&& \rstick{$\ket{0}$} \\
\lstick{$\ket{0}$} & \qwbundle{n} &&&&&&&&&&&&& \rstick{$\ket{0}$} \\
\lstick{$\ket{0}$} & \qwbundle{n} &&&&&&&&&&&&& \rstick{$\ket{0}$} \\
\lstick{$\ket{\lambda_r}$} & \qwbundle{n} &&&&&&&&&&&&& \rstick{$\ket{\lambda_r}$}
\end{quantikz}
}
    \caption{Corrected circuit for step 6 of the elliptic curve point addition circuit. The first fix involves 4 $3n$-controlled Toffoli gates and is highlighted in blue. Two of these gates are used to clear the $\ket{f_1}$ register in a rare edge case when $\ket{f_1} \neq \ket{f_2}$ and the operation is not point doubling. The other two are used to clear the $\ket{f_2}$ register in a near equivalent edge case. The Second fix involves moving the controlled ModSub and ModAdd gates to before the Equals gate and is highlighted in red. This is done so that the final values are calculated before the $\ket{f_4}$ register is cleared.}
    \label{step-6}
\end{figure*}

Step 6 contains two inconsistencies. The first surfaces when $\ket{(x,y)} = \ket{(0, 0)}$, $\ket{b} \neq\ket{0}$, and $\ket{a}=\ket{0}$ or when $\ket{(a,b)} = \ket{(0, 0)}$, $\ket{y} \neq\ket{0}$, and $\ket{x}=\ket{0}$. Because $\ket{f_1}$ represents $\ket{a}=\ket{x}$ and $0=0$, then $\ket{f_1}$ must be in the $\ket{1}$ state. It is also known that $\ket{f_2}=\ket{0}$ because $\ket{b} \neq \ket{y}$. The last gate indicates that $\ket{f_1}$ is only cleared when $\ket{(x_r, y_r)} = \ket{(0, 0)}$, however, because $\ket{b} \neq\ket{y}$ this is not true and $\ket{f_1}$ remains uncleared.

Similarly, when $\ket{(x,y)} = \ket{(0, 0)}$, $\ket{a} \neq\ket{0}$, and $\ket{b}=\ket{0}$ or when $\ket{(a,b)} = \ket{(0, 0)}$, $\ket{x} \neq \ket{0}$, and $\ket{y} = \ket{0}$. Because $\ket{f_2}$ represents $\ket{b}=\ket{-y}$ and $0=-0$, then $\ket{f_2}$ must be in the $\ket{1}$ state. We also know that $\ket{f_1}=\ket{0}$ because $\ket{a} \neq \ket{x}$. The last gate indicates that $\ket{f_2}$ is only cleared when $\ket{(x_r, y_r)} = \ket{(0, 0)}$, however, because $\ket{a} \neq\ket{x}$ this is not true and $\ket{f_2}$ remains uncleared.

The second inconsistency surfaces when $\ket{a}=\ket{x}\neq\ket{0}$ and $\ket{b}=\ket{y}=\ket{0}$; this implies that $\ket{f_1}=\ket{f_2}=\ket{1}$ and $\ket{f_4}=\ket{0}$. In this case, at the point before the equals subroutine, $\ket{(a,b)}=\ket{(x,y)}$ which flips $\ket{f_4}$ on. $\ket{f_1}=\ket{f_2}=\ket{1}$ and $\ket{f_4}=\ket{0}$, which implies that $\ket{(x,y)} = \ket{(a,\pm b)} \neq \ket{(0, 0)}$. In this case, $\ket{f_4}$ will be cleared by the equals Bloq; this being incorrect because $\ket{f_4} =\ket{0}$.

To fix the first inconsistency, I flip $\ket{f_1}$ when $\ket{(x,y)} = \ket{(0, 0)}$ and $\ket{a}=\ket{0}$ or when $\ket{(a,b)} = \ket{(0, 0)}$ and $\ket{x} = \ket{0}$; this can be seen in Fig. \ref{step-6} highlighted in blue. I do this using two multi-0-controlled gates; in each case only one of the gates will clear $\ket{f_1}$ and the final gate will not clear $\ket{f_1}$ because $\ket{(x_r,y_r)} \neq \ket{(0,0)}$. This also works when $\ket{b}=\ket{y}$ because that implies that both points are the origin and so $\ket{f_1}$ will be cleared, uncleared, and cleared again by the final gate of the circuit.

I also flip $\ket{f_2}$ when $\ket{(x,y)} = \ket{(0, 0)}$ and $\ket{b}=\ket{0}$ or when $\ket{(a,b)} = \ket{(0, 0)}$ and $\ket{y} = \ket{0}$; this can be seen in Fig. \ref{step-6} highlighted also highlighted in blue. I do this using two multi-0-controlled gates; in each case only one of the gates will clear $\ket{f_2}$ and the final gate will not clear $\ket{f_2}$ because $\ket{(x_r,y_r)} \neq \ket{(0,0)}$. This also works when $\ket{a}=\ket{x}$ because that implies that both points are the origin and so $\ket{f_2}$ will be cleared, uncleared, and cleared again by the final gate of the circuit.

To fix the second inconsistency, I move the controlled modular subtraction and controlled modular addition gates to just before the equals gate. This change can be see in Fig. \ref{step-6} and is highlighted in red. Because these gates already act before the multi-0-controlled gate at the end of the circuit, I can move them to before the equals gate without changing the clearing logic for $\ket{f_1}$ and $\ket{f_2}$. Point doubling when $\ket{b}=\ket{y}=\ket{0}$ always results in the origin point. For this reason, if I set $\ket{(x_r, y_r)}$ to the origin point before the equals gate, it will no longer flip $\ket{f_4}$.

\subsection{Resource Cost of Fixes}
Table \ref{counts} shows the recalculated subroutine counts for elliptic curve point addition alongside the Toffoli counts per subroutine. The Toffoli counts are calculated using the Bloq implementations in the Qualtran Bloq library. Fixing step 2 required one additional ancilla qubit, an additional $\textit{Equals}$ gate, and a controlled $\textit{Equals}$ gate which raises the Toffoli count by a factor of $4n$. Fixing step 5 required a $2n$-controlled Toffoli gate which raises the Toffoli count by a factor of $2n$. Step 6 required four $3n$-controlled Toffoli gates to fix part 1 which raised the Toffoli count by a factor of $12n$. No new gates were required to fix part 2; the circuit was fixed by moving the existing controlled ModSub and ModAdd gates to an earlier point in the circuit. Overall, the leading order is unchanged at $126n^2$ Toffoli gates, and the total difference is $18n-5$ additional Toffoli gates.

\begin{table}[htpb]
\centering
\caption{Revised subroutine counts for elliptic curve point addition and their corresponding Toffoli counts. Quantities that differ from Ref.~\cite{ecc-psi} are highlighted in blue. The Toffoli count of $n$-qubit modular negation, modular multiplication, and modular multiplicative inverse differ based on the cost implemented in Qualtran. All other subroutine costs are unchanged, however I include constant factors. }
    \label{counts}
\begin{tabular}{|l|r|l|}
\hline
\rowcolor[HTML]{C0C0C0} 
Subroutine                             & \multicolumn{1}{l|}{\cellcolor[HTML]{C0C0C0}Count} & Toffoli Count                    \\ \hline
$n$-controlled Toffoli                   & $\textcolor{blue}{29}$                                                 & $n-1$                              \\ \hline
$n$-qubit modular addition               & $3$                                                  & $4n-1$                             \\ \hline
$n$-qubit controlled modular addition    & $2$                                                  & $5n+1$                             \\ \hline
$n$-qubit modular subtraction            & $2$                                                  & $6n-3$                             \\ \hline
$n$-qubit controlled modular subtraction & $4$                                                  & $7n-1$                             \\ \hline
$n$-qubit modular negation               & $2$                                                  & $\textcolor{blue}{3n-3}$                             \\ \hline
$n$-qubit controlled modular negation    & $1$                                                  & $3n-2$                             \\ \hline
$n$-qubit modular doubling               & $2$                                                  & $2n+1$                             \\ \hline
$n$-qubit modular multiplication         & $10$                                                 & $\textcolor{blue}{2.25n^2+7.25n-1}$ \\ \hline
$n$-qubit modular multiplicative inverse & $4$                                                  & $\textcolor{blue}{26n^2+9n-1}$      \\ \hline
$n$-qubit equals                         & $\textcolor{blue}{6}$                                                  & $n-1$                              \\ \hline
$n$-qubit controlled equals              & $\textcolor{blue}{1}$                                                  & $\textcolor{blue}{3n}$                               \\ \hline
\end{tabular}
\end{table}

\section{Conclusion}
By applying these corrections to the elliptic curve point addition circuit, an exact circuit is constructed for the operation that works in all situations and correctly returns all ancilla qubits used to their original state while leaving the leading-order gate cost unchanged. The corrections contribute a total of $18n-5$ Toffoli gates. The work done here highlights the importance of verifying the correctness of quantum circuits when possible. Libraries and frameworks such as Qualtran can provide an efficient way to unit test classically simulatable quantum circuits.

\section{Code Repository}
The Qualtran code is available at \url{https://github.com/quantumlib/Qualtran} in the bloqs/cryptography/ecc subdirectory. The modifications to the circuit were submitted to the repository in \url{https://github.com/quantumlib/Qualtran/pull/1489} and \url{https://github.com/quantumlib/Qualtran/pull/1666}.

\section*{Acknowledgment}
I thank Matthew Harrigan, Noureldin Yosri, and Anurudh Peduri for their work implementing key functionality in the library, offering low-level arithmetic circuits required to realize the high-level circuits described in this work, and their diligent code reviews. I also thank Sam Pallister and Angus Kan
for their reviews which helped to correct an error in the circuit designs.

\bibliographystyle{unsrt}
\bibliography{reference}
\end{document}